\documentclass[runningheads,a4paper]{llncs}

\let\llncssubparagraph\subparagraph
\let\subparagraph\paragraph
\usepackage[compact]{titlesec}
\let\subparagraph\llncssubparagraph

\usepackage{amssymb}
\usepackage{titlesec}
\usepackage{graphicx}
\usepackage{multirow}
\usepackage{booktabs}
\usepackage{adjustbox}
\usepackage{rotating}
\usepackage{url}

\urldef{\mailsa}\path|{ugur, manas, usha, prasad,amit}@knoesis.org|
\urldef{\mailsb}\path|{kursuncu, budak}@uga.edu|

\titleformat{\subsubsection} {\normalfont\normalsize\bf}{\thesubsubsection}{1em}{} \titlespacing*{\subsubsection}{0pt}{3.25ex plus 1ex minus .2ex}{0ex plus .2ex}

\begin{document}

\mainmatter  

\title{Predictive Analysis on Twitter: Techniques and Applications}

\titlerunning{Predictive Analysis on Twitter: Techniques and Applications}

%
%
\author{Ugur Kursuncu\textsuperscript{1,2}
\and  Manas Gaur\textsuperscript{1}\and Usha Lokala\textsuperscript{1}\and Krishnaprasad Thirunarayan\textsuperscript{1}\and\\
Amit Sheth\textsuperscript{1}\and I. Budak Arpinar\textsuperscript{2}}
%
\authorrunning{Kursuncu, Gaur, Lokala, Thirunarayan, Sheth and Arpinar}

\institute{\textsuperscript{1}Kno.e.sis Center, Wright State University, Dayton, OH, USA\\
\textsuperscript{2}Department of Computer Science, The University of Georgia, Athens, GA, USA\\
\mailsa\\
\mailsb\\}

%
%

\toctitle{Lecture Notes in Computer Science}
\tocauthor{Authors' Instructions}
\maketitle

\begin{abstract}

Predictive analysis of social media data has attracted considerable attention from the research community as well as the business world because of the essential and actionable information it can provide. Over the years, extensive experimentation and analysis for insights have been carried out using Twitter data in various domains such as healthcare, public health, politics, social sciences, and demographics. In this chapter, we discuss techniques, approaches and state-of-the-art applications of predictive analysis of Twitter data. Specifically, we present fine-grained analysis involving aspects such as sentiment, emotion, and the use of domain knowledge in the  coarse-grained analysis of Twitter data for making decisions and taking actions, and relate a few success stories.
\end{abstract}

\section{Introduction}
With the growing popularity of social media and networking platforms as an important communication and sharing media, they have significantly contributed to the decision making process in various domains. In the last decade, Twitter has become a significant source of user-generated data. The number of monthly active users was 330 million as of third quarter of 2017, and the number of daily active users was 157 million as of second quarter of 2017.  Moreover, nearly 500 million tweets per day are shared on Twitter. Accordingly, significant technical advancements have been made to process and analyze social media data using techniques from different fields such as machine learning, natural language processing, statistics, and semantic web. This amalgamation and interplay of multiple techniques within a common framework have provided feature-rich analytical tools \cite{Purohit2013,Davis2016a}, leading to valid, reliable and robust solutions.

Twitter provides multimodal data containing text, images, and videos, along with contextual and social metadata such as temporal and spatial information, and information about user connectivity and interactions. This rich user-generated data plays a significant role in gleaning aggregated signals from the content and making sense of public opinions and reactions to contemporary issues. Twitter data can be used for predictive analysis in many application areas, ranging from personal and social to public health and politics. Predictive analytics on Twitter data comprises a collection of techniques to extract information and patterns from data, and predict trends, future events, and actions based on the historical data.

\begin{figure}
		\centering
		\includegraphics[width=12.0cm]{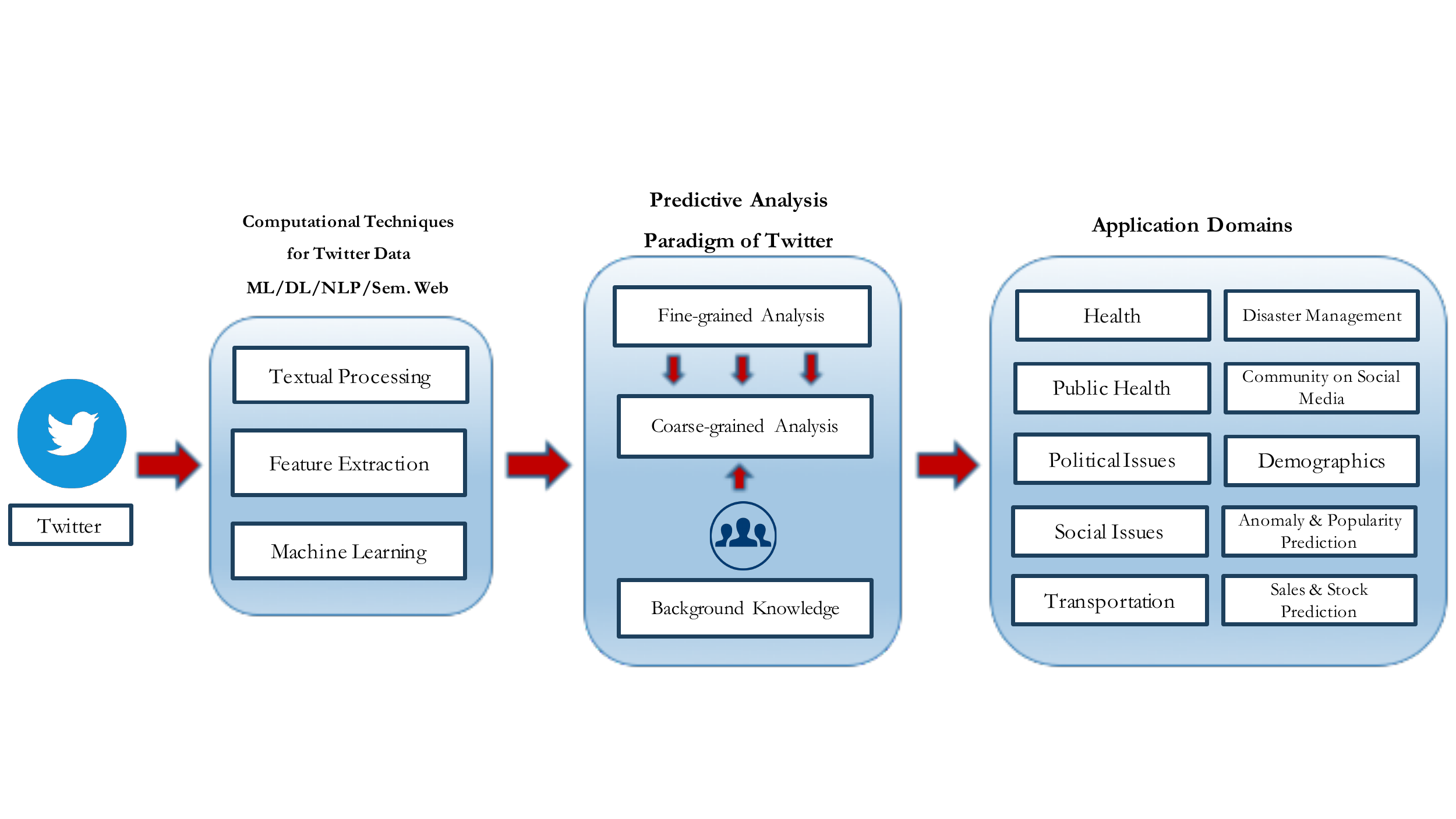}
		\caption{Overview of Predictive Analysis on Twitter Data.}
		\label{fig:example}
		\end{figure}

Gaining insights and improving situational awareness on issues that matter to the public are challenging tasks, and social media can be harnessed for a better understanding of the pulse of the populace. Accordingly, state-of-the-art applications, such as Twitris \cite{Sheth2018a} and OSoMe \cite{Davis2016a}, have been developed to process and analyze big social media data in real time. Regarding availability and popularity, Twitter data is more common than data from web forums and Reddit\footnote{\url{https://goo.gl/Jo1h9U}}. It is a rich source of user behavior and opinions. Although analytical approaches have been developed to process Twitter data, a systematic framework to efficiently monitor and predict the outcome of events has not been presented. Such a framework should account for the granularity of the analysis over a variety of domains such as public health, social science, and politics, and it has been shown in Figure \ref{fig:example}.

We discuss a predictive analysis paradigm for Twitter data considering prediction as a process based on different levels of granularity. This paradigm contains two levels of analysis: \textit{fine-grained} and \textit{coarse-grained}. We conduct fine-grained analysis to make tweet-level predictions on domain independent aspects such as sentiment, topics, and emotions. On the other hand, we perform coarse-grained analysis to predict the outcome of a real-world event, by aggregating and combining fine-grained predictions. In the case of fine-grained prediction, a predictive model is built by analyzing social media data, and prediction is made through the application of the model to previously unseen data. Aggregation and combination of these predictions are made from individual tweets form signals that can be used for coarse-grained predictive analysis. In essence,  low-level signals from tweets, such as sentiment, emotions, volume, topics of interest, location and timeframe, are used to make high-level predictions regarding real-world events and issues. 

In this chapter, we describe use of Twitter data for predictive analysis, with applications to several different domains. In Section 2, we discuss both processing and analytic techniques for handling Twitter data and provide details of feature extraction as well as machine learning algorithms. In Section 3, we explain a predictive analysis paradigm for Twitter that comprises two levels: fine-grained and coarse-grained. We also provide use cases, based on real-world events, of how coarse-grained predictions can be made by deriving more profound insights about a situation from social media using signals extracted through fine-grained predictions. We also describe common domain-independent building blocks that can serve as the foundation for domain-specific predictive applications. In Section 4, we give further details on specific state-of-the-art applications of Twitter analytics that have been developed for different domains, such as public health, social and political issues. In Section 5, we conclude with a discussion of the impact of social media on the evolvement of real-world events and actions, challenges to overcome, for broader coverage and more reliable prediction. We also provide a comparative table relating techniques used with corresponding applications. 

\section{Language Understanding of Tweets}
Novel processing and analysis techniques are required to understand and derive reliable insights to predict trends and future events from Twitter data due to their unique nature -- it contains slangs, unconventional abbreviations and grammatical errors as a matter of course. Moreover, due to the evolving nature of many events, may it be political, sports, or disaster-related, collecting relevant information as the event unfolds is crucial  \cite{Penuel2011,Malilay2014}. Overcoming the challenges posed by the volume, velocity, and variety of incoming social, big data is non-trivial \cite{Wang2012}. Sole keyword-based crawling suffers from low precision as well as low recall. For instance, obtaining tweets related to marijuana legislation \cite{Lamy2017} using its street name  “spice” pulls irrelevant content about “pumpkin spice latte” and “spice” in food. To improve recall without sacrificing precision, Sheth et al. \cite{Sheth2016} provided a solution for adapting and enhancing filtering strategies that (a) obtains customized tweet streams containing topics of user interest \cite{Kapanipathi2011} by constructing a hierarchical knowledge base by analyzing each user’s tweets and profile information \cite{Kapanipathi2014}, (b) selects and employs a domain-specific knowledge graph (e.g., using the Drug Abuse Ontology for opioid related analysis \cite{Cameron2013}) for focus, and (c) reuses a broad knowledge graph such as DBPedia for coverage and generality.
In Twitter data analysis, the processing phase includes natural language processing using techniques such as TF-IDF, word2vec, stemming, lemmatization, eliminating words with a rare occurrence, and tokenizing. On the other hand, some of the commonly used techniques, such as removal of stop-words, have proven ineffective.  Saif \cite{Saif2017a} has compared six different stop words identification methods over six different Twitter datasets using two well-known supervised machine learning methods and assessed the impact of removing stop words by observing fluctuations in the level of data sparsity, the size of the classifier’s feature space and the classifier performance. Saif concludes that in most cases that removing stop words from tweets has a negative impact on the classification performance. 
\subsection{Unique Nature of Tweets}
Twitter’s limit on the number of characters in a message encourages the use of unconventional abbreviations, misspellings, grammatical errors and slang terms. For instance, since a tweet was limited to 140 characters (until recent doubling to 280 character in December 2017), different sets of techniques and metadata have been considered to identify the best features to optimize the overall performance of the model being built. Due to the heterogeneous nature of the Twitter content, one can develop a variety of features \cite{Wijeratne2017b} ranging from textual, linguistic, visual, semantic, network-oriented, to those based on the tweet and user metadata. Further, to handle tweet’s textual data, the extracted features, techniques and tools \cite{Sheth2018a,Gimpel2011,Wagner2013a} have been customized to exploit as well as being robust concerning misspellings, abbreviations, and slangs.  Gimpel et al. \cite{Gimpel2011} addressed this problem in the context of part-of-speech (PoS) tagging, by developing a new tagset along with features specific to tweets, and reported 89\% accuracy as opposed to Stanford tagger with 85\% accuracy.

Tweets also include hashtags, URLs, emoticons, mentions, and emoji in their content. As these components contribute to the meaning of a tweet, it is imperative that we incorporate them in the analysis, on a par with textual content. 

\textbf{Hashtags} are meant to help in categorizing tweet's topics. They are frequently used to collect and filter data as well as for sentiment \cite{Wang2011,Davidov2010,Kouloumpis2011a}, emotion \cite{Wang2012}, and topical analysis \cite{Romero2011,Morstatter2013}. Wang et al. \cite{Wang2011} used hashtags in their topical hashtag level sentiment analysis incorporating co-occurrence and literal meaning of hashtags as features in a graph-based model and reported better results compared to a sentiment analysis approach at the tweet level. In emotion analysis, Wang et al. \cite{Wang2012} collected about 2.5 million tweets that contain emotion-related hashtags such as \#excited, \#happy, and \#annoyed, and used them as the self-labeled training set for developing a high accuracy, supervised emotion classifier.

\textbf{URL} presence in a tweet is usually indicative the content being an index for a longer explanatory story pointed to by the URL. Researchers found URLs in a tweet to be discriminative in various studies such as sentiment analysis \cite{Go2009,Agarwal2011}, popularity prediction \cite{Suh2010,Naveed2011a}, spam detection \cite{Thomas2011a}. They reported that the feature for URL presence in a tweet appeared as a top feature or has a substantial contribution to the accuracy of the model.

\textbf{Emoticons} (e.g., :), $<3$) have been exploited by Liu et al. \cite{Liu2012} in their Twitter sentiment analysis study, such as by interpreting “:)” as conveying positive sentiment and “:(“ as conveying the negative sentiment. They used all tweets containing those emoticons as self-labeled training set and integrated them with the manually labeled training set \cite{Zhai2004}. They have achieved significant improvement over the model trained with only manually labeled data. Go et al. \cite{Go2009}, and other researchers \cite{Boia2013,Pak2010} conducted sentiment analysis on Twitter in 2009, and they found that they were able to achieve a better accuracy using models trained with emoticon data. 

\textbf{Emoji} is a pictorial representation of facial expressions, places, food and many other objects, being used very often on social media to express opinions and emotions on contemporary issues of contentions and discussions. The use of emoji is similar to emoticon since they both provide a shorter means of expression of an idea and thought. The difference is that an emoji use a small image for the representation as opposed to emoticon that uses a sequence of characters. Kelly et al. \cite{Kelly2015} studied the use of emoji in different contexts by conducting interviews and found that the use of emoji goes beyond the context that the designer intended. Novak et al. \cite{Novak2015} created an emoji sentiment lexicon analyzing the sentiment properties of emojis, and they pointed that the emoji sentiment lexicon can be used along with the lexicon of sentiment-bearing words to train a sentiment classifier. On the other hand, Miller et al. \cite{Miller2016} found that the emoji provided by different platforms are not interpreted similarly. Wijeratne et al. \cite{Wijeratne2017a} gathered possible meanings of 2,389 emojis in a dataset called EmojiNet, providing a set of words (e.g., smile), its POS tag (e.g., verb), and its definition, that is called its “sense.” It associates 12,904 sense labels with 2,389 emojis, addressing the problem of platform-specific meanings by identifying 40 most confused emoji to a dataset.

\subsection{Metadata for Tweet and User}
There are mainly two types of metadata in a tweet object, namely, tweet metadata\footnote{\url{Tweet Object. https://developer.twitter.com/en/docs/tweets/data-dictionary/overview/tweet-object}} and user metadata\footnote{\url{User Object. https://developer.twitter.com/en/docs/tweets/data-dictionary/overview/user-object}}. Tweet metadata contains temporal and spatial information along with user interactions and other information such as replies and language. On the other hand, user metadata contains information pertaining to the user that authored the tweet, such as screen-name and description. Some of the available metadata are described below.
\subsubsection{Tweet Metadata} 
\textbf{createdAt}: This field contains the information on when the tweet was created, which is especially important when a time series analysis is being done \cite{Varol2017b}.\\
\textbf{favoriteCount}: The users on Twitter can like a tweet, and this is one way of interacting with the platform. The number of likes for a tweet has been used as a feature in various applications that includes trend detection \cite{Varol2017b}, identification of influence and popularity.\\
\textbf{inReplyToScreenName}: If this field of the tweet object is not null, it is a reply to another tweet, and this field will hold the username of the user that authored the other tweet. This information is valuable, especially to predict the engagement of the audience over an issue that tweets relate to, and to find influential users.\\
\textbf{geoLocation}: the Twitter platform has a feature that can attach the users’ geolocation to the tweet, but this is up to the users to make it publically available. Most of the users prefer not to share their geolocation.\\
\textbf{retweet\_count}: Twitter allows users to repost a tweet by retweeting to their audience, and the original tweet holds this field to keep how many times this tweet has been retweeted. This information is useful to incorporate the prediction of popularity and trending topics.

\subsubsection{User Metadata}
\textbf{description}: This field holds the description of the account. As this metadata carries information on characteristics of the user, it is mostly used in user classification.\\
\textbf{followers\_count}: This field holds the number of followers the user has, and as it is changeable information over time, the information located in a specific tweet may not be up to date.\\
\textbf{friends\_count}: Twitter calls the accounts that a user follows as "friends," but it is also known as "followees." The numbers of followers and followees are used to determine the popularity of user and topics.\\
\textbf{statuses\_count}: Twitter also calls tweets as “status,” and in this case, status count refers to the number of tweets that a user has posted. 

\subsection{Network and Statistical Features }
The users interact on the social networking platform Twitter with each other through follows, replies, retweets, likes, quotes, and mentions. Centrality metrics have been developed to compute and reveal users’ position and their importance based on their connections in their network. These centrality measures can help identify influential users. These metrics include in-degree, out-degree, closeness, betweenness, PageRank and eigenvector centrality. Closeness centrality is defined by Freeman \cite{Freeman1978} as the sum of distances from all other nodes, where the distance from a node to another is defined as the length (in links) of the shortest path from one to the other. The smaller the closeness centrality value, the more central the node. Betweenness \cite{Freeman1977a} measures the connectivity of a node by computing the number of shortest paths which pass through the node. This aspect makes this node, a user in a Twitter social network, an essential part of the network as it controls the flow of information in the network. Therefore, removing this node would disconnect the network. EigenVector \cite{Bonacich1987,Lawyer2015} metric measures the importance of a node based on the importance of its connections within the network. Therefore, the more critical connections a node gets, the more critical the node becomes. These metrics were used in a user classification application as features by Wagner et al. \cite{Wagner2013a} because of the intuition that similar users would have similar network connectivity characteristics. 

Statistical features such as min, max, median, mean, average, standard deviation, skewness, kurtosis, and entropy can be computed for several data attributes \cite{Varol2017b}. Machine learning determines a subset of these features that have the discriminative power necessary for particular applications and domains, especially for predicting user behaviors and user types \cite{Pennacchiotti2011}. For instance, \cite{Varol2017b} extracted statistical features of a user, tweet, network. The statistical analysis was done over attributes such as sender’s follower count, originator’s followee count, the time between two consecutive tweets, and the number of hashtags in a tweet. They conducted a time series analysis to predict if a trending meme is organic or promoted by a group. On the other hand, \cite{Pennacchiotti2011} utilized statistical features to predict the type of users on social media based on their political leanings, ethnicity, and affinity for a particular business. As they classified users, they computed statistical characteristics of tweeting behavior of users such as average number of messages per day, average number of hashtags and URLs per tweet, average number and standard deviation of tweets per day.

\subsection{Machine Learning and Word Embeddings}

Machine learning algorithms play a crucial role in the predictive analysis for modeling relationships between features.  It is well-known that there is no universal optimal algorithm for classification or regression task, and in fact requires us to tailor the algorithm to the structure of the data and the domain of discourse. Recent survey papers \cite{Irfan2004,Nassirtoussi2014,Franch2013,Bravo-Marquez2012} and our comparative analysis ({see Table \ref{table_1}}) of related influential studies show what algorithms we found to perform well for various applications. As can be seen, this covers a wide variety -- Random Forest, Naive Bayes, Support Vector Machine, Artificial Neural Networks, ARIMA and Logistic Regression. 

Furthermore, deep learning (a.k.a advanced machine learning) enhanced the performance of learning applications. Deep learning is a strategy to minimize the human effort without compromising performance. It is because of the ability of deep neural networks to learn complex representations from data at each layer, where it mimics learning in the brain by abstraction\footnote{How do Neural networks mimic the human brain? \url{    https://www.marshall.usc.edu/blog/how-do-neural-networks-mimic-human-brain}}. The presence of big data, GPU, and sufficiently large labeled/unlabeled datasets improve its efficacy. We discuss some of the applications that make use of deep learning for prediction task on social media in section 4.

Textual data processing benefits from the lexico-semantic representation of content. TF-IDF \cite{Hong2011}, Latent/Hierarchical Dirichlet Allocation(LDA/HDA) \cite{Sokolova2016}, Latent Semantic Analysis (LSA) \cite{Dumais2008} and Latent Semantic Indexing have been utilized in prior studies for deriving textual feature representations. In a recent paper \cite{Mikolov2013}, they put forward a word embedding approach called Word2Vec that generates a numerical vector representation of a word that captures its contextual meaning incorporating its nearby words in a sentence. Training the word embedding model on a problem-specific corpus is essential for high-quality domain-specific applications, since the neighborhood set of words for an input term impacts its word embedding. For instance, pre-trained models of word2vec on news corpora generate poor word embeddings over a Twitter corpus. Wijeratne et al. \cite{Wijeratne2016a} used word embeddings to further enhance the prediction of gang members on Twitter by training their model on a problem-specific corpus. 

\subsection{Multi-modality on Twitter}

Visual elements such as images and videos are often used on social media platforms. While users can attach images and videos to their tweets, they can also upload a profile image and a header image. Since the latter images are mostly related to the user’s characteristics, personality, interest or a personal preference, these images are mostly used for classification of account type (e.g., media, celebrity, company), detection of user groups \cite{Balasuriya2016,Wijeratne2016a} and identification of demographic characteristics (e.g., gender, age) \cite{Sakaki2014}. Balasuriya et al. \cite{Balasuriya2016} used the profile image of users in their feature set for finding street gang members on Twitter since gang members usually set their profile image in a particular way to intimidate other people and members of rival gangs. They retrieved a set of 20 words and phrases for each image through the Clarifai\footnote{\url{https://www.clarifai.com}} web service to be used as features. As image processing is costly regarding time and computational resources required for training a model to retrieve information from images, it is usually preferred to use off-the-shelf web services that provide cheaper, yet effective alternative, for scalable social media analytics.

\section{Prediction on Twitter Data}

Gaining understanding about and predicting an event’s outcome and its evolvement over time using social media, requires incorporation of analysis of data that may differ in granularity and variety.  As tools \cite{Gimpel2011,Bontcheva2013a} are developed and customized for Twitter, its dynamic environment requires human involvement in many aspects. For instance, verification of a classification process \cite{Mitra2016} and annotation of a training dataset \cite{Wijeratne2017a,DeChoudhury2013a,Lewenberg2015} are essential in the predictive analysis that can benefit from human expert guidance in creating ground truth dataset. Social media analysis in the context of complex and dynamic domains \cite{Wang2012b,Ebrahimi2017,Vieweg2010a} is challenging. Our approach to overcoming this challenge and dealing with a variety of domains is to customize domain independent building blocks to derive low-level/fine-grained signals from individual tweets. Then, we aggregate and combine these signals to predict high-level/coarse-grained domain-specific outcomes and actions with a human in the loop. 

	    \begin{figure}
		\centering
		\includegraphics[width=12.0cm]{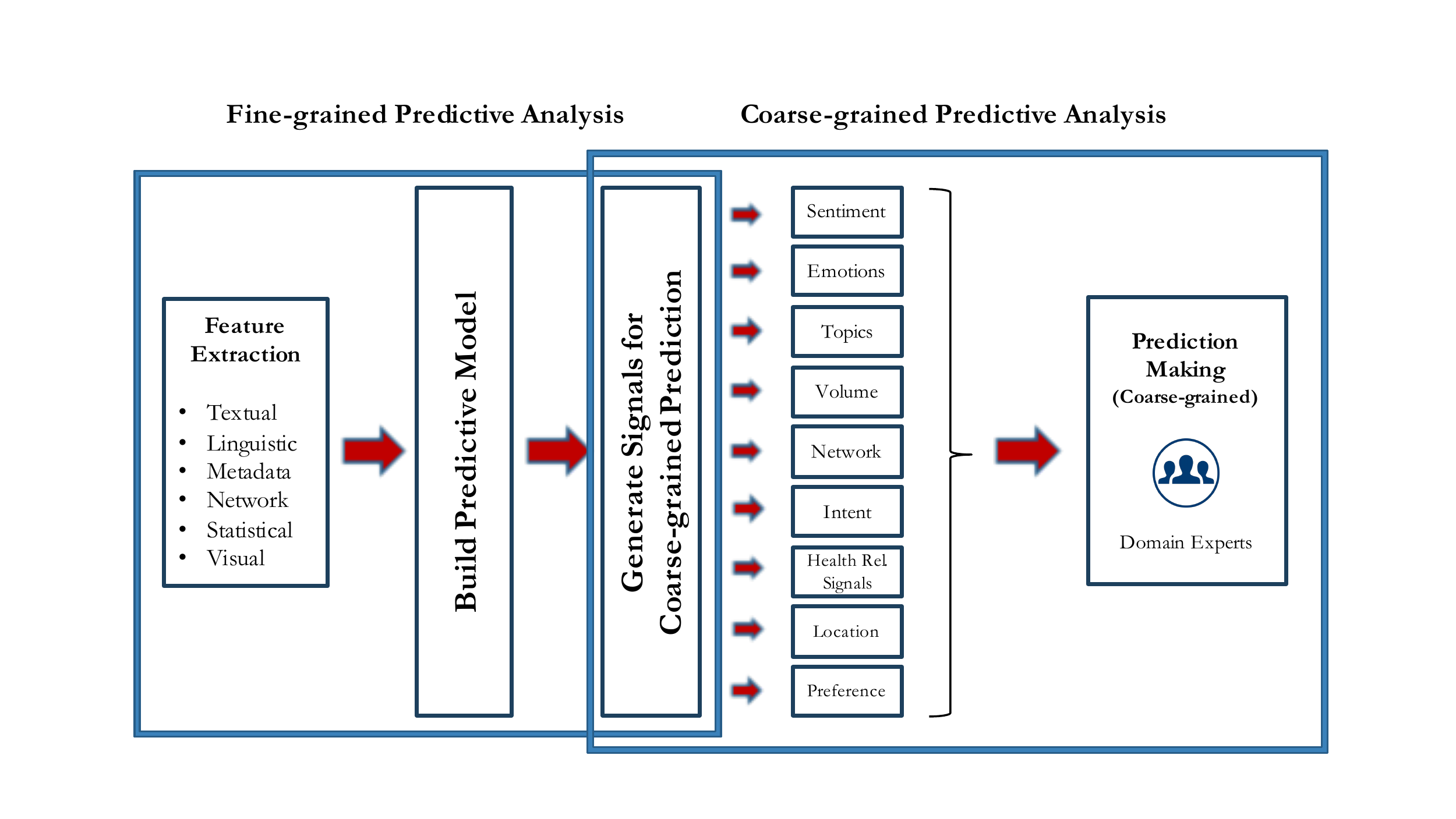}
		\caption{Two level Predictive Analysis Paradigm for Twitter.}
		\label{fig:predParad}
		\end{figure}

\subsection{A Predictive Analysis Paradigm for Twitter}

We consider predictive analysis on Twitter data as a two-phase approach: The first phase is fine-grained predictive analysis and the second phase is coarse-grained predictive analysis. An illustration of this paradigm is depicted in figure \ref{fig:predParad}. The fine-grained analysis is a tweet-level prediction for individual signals, such as sentiment and emotions, about an event that is being monitored. This low-level prediction is made by building a predictive model that employs feature engineering and machine learning algorithms. Aggregating the tweet-level predictions for a specific time frame and location generates signals. For instance, a predictive model for sentiment predicts the sentiment of each tweet about an event in question as negative (-1) neutral (0) or positive (+1), and we produce a signal between -1 and +1 for a particular location and time frame. A collection of such signals (e.g., emotions, topics)  helps domain experts form insights while monitoring or predicting the outcome of an event, in their higher level analysis. Extraction of these signals is discussed further in subsequent section. 

Coarse-grained analysis is a higher level prediction involving outcomes and trends of a real-world event, such as elections\cite{Chen2012}, social movements\cite{DeChoudhury2016} and disaster coordination\cite{Purohit2013a,Purohit2014,Purohit2014a,ShreyanshBhatt2014}. In this case, we gather the signals which we generated from the fine-grained predictions and make a judgment call for the outcome by making sense of these signals in the context of the event and the related domain. Sentiment, emotions, volume, topics, and interactions between Twitter users can be considered as signals, while the importance and informativeness of each of these parameters may vary depending on the event and its domain. For instance, gauging the sentiment of a populace towards an electoral candidate would be very significant to predict the outcome of an election \cite{Ebrahimi2017}, but the same kind of information may not be as critical in the context of disaster management because, in the latter case, the sentiment may be largely negative. Further, for reliable decision making, the sentiment must be interpreted in a broader context. Predominantly positive sentiment towards democratic candidates in California is not as significant as that in Ohio. Similarly, the context provided by county demographics may be crucial in generalizing, predicting, and determining the outcome of an election. Moreover, temporal and spatial context plays an important role to understand the ongoing events better and obtain more profound insights. In US presidential elections, some states, called the “swing states” (as the electorate’s choice has changed between Republican and Democratic candidates through the previous elections in these states), typically determine the eventual outcome of US elections. Therefore, narrowing down the analysis to the state level and gathering signals from these particular states would meaningfully contribute to the prediction of the outcome of the Presidential election and the future direction of the country.

In general, prediction analytics requires domain-specific labeled datasets created with the assistance of domain experts, and customization of feature space, classification algorithm, and evaluation. Real world events have a dynamic nature in which critical happenings may change the course of discussions on social media.  For example, breaking news about a candidate in an election may change the vibe in echo chambers of Twitter; thus, affecting the public opinion in one or another direction. For this reason, it is imperative to conduct the analysis accounting for essential milestone events happening during the process. Therefore, the analysis of such events would require an active learning paradigm that incorporates evolving domain knowledge in real-time.

\subsection{Use Cases for Coarse-grained Prediction}
Coarse-grained prediction requires taking into account many signals, and evaluating them concerning both present and historical context that varies with location and time frame. Importance of the signals in some domains and their related events may vary, and sole use of these signals would not be sufficient to make a reliable judgment call, although these signals are essential parameters in a real-world event context. For instance, an election usually whips up discussions on various sub-topics, such as unemployment, foreign policy; and necessitates proper cultivation of a diverse variety of signals following contextual knowledge of the domain \cite{Ebrahimi2017}. We provide two use cases in this subsection to illustrate how a coarse-grained or high-level predictive analysis can be conducted.
\subsubsection{US 2016 Presidential Election}
During the 2016 US Presidential elections where “swing states” played a key role in determining the outcome, many polling agencies failed to predict it accurately\footnote{\url{http://www.pewresearch.org/fact-tank/2016/11/09/why-2016-election-polls-missed-their-mark/}}\footnote{\url{https://goo.gl/mFtzvb}}. On the other hand, researchers\footnote{\url{https://goo.gl/AJVpKf}} conducted a real-time predictive analysis using a social media analytics platform \cite{Sheth2018a}, making the prediction accurately before the official outcome was announced, by analyzing the state-level signals, such as from Florida and Ohio. Temporal aspect was also important in this use case to explain the evolution of the public opinion based on milestone events over the period of the election, as well as the election day because people tend to express, who they voted in the same day. They analyzed 60 million tweets by looking at the sentiment, emotions, volume, and topics narrowing down their analysis to state-level. On the election day, they focused on specific states such as Florida, which, before the election day, they predicted would be a pathway for Donald Trump to win the election\footnote{\url{https://goo.gl/sh7WNr}}. In their analysis of Florida, volume and positive emotion (joy) for Trump was higher, whereas positive sentiment for Clinton was higher, eliciting report\footnote{\url{https://goo.gl/iCqzk3}} such as “limited to professed votes from Florida until 1pm is not looking in her favor”. Later in the day, the volume of tweets for Trump increased to 75\% of all tweets based on the hashtag "\#ivoted". Particularly in critical states of Florida, North Carolina, and Michigan, volume and positive emotions for Trump were significantly higher than for Clinton, although the sentiment was countering the overall signal. They made the call that the winner of Presidency and Congress as Donald Trump and the GOP respectively. While conducting this analysis \cite{Ebrahimi2017}, they noticed that the predictive model that they have built for sentiment signal was not successful due to the dynamic nature of the election with changing topics in conversations. A similar analysis was made for UK Brexit polls in 2012 by the same researchers, correctly predicting the outcome utilizing the volume and sentiment signals\footnote{\url{https://goo.gl/i2Ztm6}}\footnote{\url{https://goo.gl/dFCGL9}}\footnote{\url{https://goo.gl/2EhSma}}.

\subsubsection{US Gun Reform Debate 2018}
Researchers\footnote{\url{http://blog.knoesis.org/2018/04/debate-on-social-media-for-gun-policy.html}} monitored gun reform discussions on Twitter to predict the public support using the Twitris platform after the tragic shooting at a high school in Parkland, Florida, in February 2018. The public started demanding a gun control policy reform, and it has attracted the attention of legislative and executive branches of both state and federal governments. As polls measured the public opinion\footnote{\url{http://time.com/5180006/gun-control-support-has-surged-to-its-highest-level-in-25-years/}}, researchers reported that the public support for a gun reform on social media was increasing over time since the Parkland shooting, confirming the overall outcome of these polls. They observed that reactions from public on social media in terms of the volume, sentiment, emotions and topics of interest, are strongly aligned with the milestone events related to this issue such as (i) POTUS’ (President of the United States) meeting with families of the victims on February 21, (ii) CPAC(Conservative Political Action Conference) between February 22 and 24, (iii) POTUS’ meeting with lawmakers on February 28 expressing strong support for a gun control policy change. These events significantly affected the public opinion on social media based on the aforementioned signals.

At the beginning of the gun reform discussions on social media, sentiment for pro-gun reform tweets was strong whereas the sentiment for anti-gun reform was relatively weak. However, the CPAC meeting changed the climate on social media, and it significantly boosted the momentum of anti-gun reform tweets, especially after the NRA (National Rifle Association) CEO Wayne LaPierre’s speech in the morning of February 22\footnote{\url{https://goo.gl/kgbqWC}}. Overall the volume of tweets for pro-gun reform was mostly higher than the anti-gun overhaul, except between February 22 and February 24\footnote{\url{https://goo.gl/LMFu3B}}, which covers the CPAC meeting where NRA CEO, VP Pence, and POTUS gave speeches. It surged the volume, positive sentiment and emotions in anti-gun reform posts radically, and those parameters for pro-gun reform posts dropped in the same manner. Effect of the meeting lasted a few days, and boycott calls for NRA and NRA’s sponsors started to pick up in the meantime. After the meeting, sentiment for pro-gun reform tweets increased consistently, and the emotions expressed in pro-gun reform tweets became more intensified.

Emotions in anti-gun reform tweets were intense especially during and after the CPAC meeting, but later emotions in pro-gun reform tweets took over. Especially volume, positive sentiment, and emotions were overwhelmingly high right after the POTUS meeting with lawmakers on Wednesday, February 28, expressing his support for a gun policy reform.

Furthermore, some of the most popular topics that users were discussing in their tweets included “midterm elections”, “parkland students”, “boycott the nra”, “stupid idea” and “individual freedoms”, where pro-gun reform arguments were expressed more frequently. The topic of “midterm elections” being one of the most popular topics on social media in gun reform discussions, also suggests that politicians from both Democrats and Republicans sensed the likely effect of this public opinion change on the midterm elections on November 2018. They have concluded in their predictive analysis that the public support for gun reform was significantly higher based on the signals they observed in the context of related events.

\subsection{Extraction of Signals}
We make predictions for the outcome of real-world events based on the insights we collect from big social data, and these insights are extracted as various signals such as sentiment, emotions, volume, and topics. The sentiment is a qualitative summary of opinions on a particular issue, and sentiment analysis techniques are utilized to extract such information computationally. The emotional analysis provides another stream of qualitative summary that is expressed by users about a particular event. The volume of tweets is an important signal about the engagement of the public in an event or an issue of consequence. Topical analysis is a process that extracts topics that contain particular themes in the domain of interest. We can produce and make use of more specific signals depending on the domain such as preference, intent, and symptoms. The signals described below are commonly used parameters in higher level prediction tasks, and we describe related state-of-the-art applications and their technical details in the following.

\subsubsection{Sentiment Analysis}
Sentiment is one of the essential signals that can be used to measure the public opinion about an issue. As users on Twitter express their opinions freely, sentiment analysis of tweets attracted the attention of many researchers.  Their approaches differ regarding the feature set, machine learning algorithm, and text processing techniques. Considering feature set, \cite{Kouloumpis2011a} used n-grams, POS-tags, emoticons, hashtags and subjectivity lexicon for sentiment analysis. For machine learning, Naive Bayes, SVM, and Conditional Random Fields (CRF) have been employed, and Naive Bayes has shown good performance \cite{Pak2010}. Also, text processing techniques like stopwords removal, word-pattern identification, and punctuation removal have shown to improve sentiment analysis in \cite{Davidov2010}. Nguyen et al. \cite{Nguyen2012} used time series analysis to be able to predict the public opinion so that the stakeholders on a stock market can react or pro-act against the public opinion by “posting balanced messages to revert the public opinion” based on the measurement that they performed using social media. Their objective was to use the sentiment change over time by identifying key features that contribute to this change. They measured the sentiment change regarding the fraction of positive tweets. They employed SVM, logistic regression and decision tree, and found that SVM and logistic regression provided similar results outperforming the decision tree. They modeled the sentiment change overall twitter data and achieved around 73\% F-score on sentiment prediction using time series analysis. \cite{Stojanovski2016} employs a deep learning approach combining convolutional and gated recurrent neural network (CGRNN) for a diverse representation of tweets for sentiment analysis. Such a system was trained on GloVe word embedding created on a crawled dataset.  The system was ranked among the top 10, evaluated using average F1 score, average recall, mean absolute error (MAE), Kullback-Leibler divergence (KLD), and EMD score \cite{Esuli2015} for SemEval-2016 sub-tasks B, C, D, E. Exclusion of hand-crafted features and improved performance on SemEval 2016 shows the potency of the approach.

\subsubsection{Emotion Analysis}
Identification of emotions in tweets can provide valuable information about the public opinion on an issue. Wang et al. \cite{Wang2012} predicted seven categorical emotions from the content of tweets using 131 emotion hashtags and utilizing the features such as n-grams, emotion lexicon words, part-of-speech tags, and n-gram positions. They used two machine learning algorithms: LIBLINEAR  and Multinomial Naive Bayes. In a similar study, Lewenberg et al. \cite{Lewenberg2015} examined the relationship between the emotions that users express and their perceived areas of interest, based on a sample of users. They used Ekman’s emotion categories and crowdsourced the task of examining the users and their tweet’s content to determine the emotions as well as their interest areas. They created a tweet-emotion dataset consisting of over 50,000 labeled tweet-emotion pairs, then trained a logistic regression model to classify the emotions in tweets according to emotion categories, using textual, linguistic and tweet metadata features. The model predicted a user’s emotion score for each emotion category, and they determined the user's interest in areas such as sports, movies, technical computing, politics, news, economics, science, arts, health, and religion.

\subsubsection{Topical Analysis}
Topical analysis is one of the essential strategies under the umbrella of information extraction techniques that capture semantically relevant topics from the social media content \cite{Griffiths2007}. Extraction of topics in the context of social media analysis helps understand the subtopics associated with an event or issue and what aspects of the issue have attracted the most attention from the public. As discussed in use cases for elections and gun reform debate, it is imperative to have the topics extracted from tweets for a better understanding of the underlying dynamics of relevant discussions.  Chen et al. \cite{Chen2012a} associated topics of interest with their relative sentiment to monitor the change in sentiments on the extracted topics. Furthermore, utilizing the extracted topics as features for a supervised model improved the performance of the classification task in \cite{Hong2010}. In \cite{Zhao2011}, researchers assessed quality of topics using coherence analysis, context-sensitive topical PageRank based ranking and probabilistic scoring function. This approach was used in a crime prediction application \cite{Wang2012d}. 

\subsubsection{Engagement Analysis}
The volume is the size of the dataset that has been collected and indicates the user engagement on an event being monitored. In general, the larger the dataset, the better is the accuracy and consistency of a predictive model because it minimizes the possibility of bias. Engagement analysis enables human experts to improve their confidence in the learned representations/patterns for an accurate high-level prediction. However, while maintaining the sufficient size of the dataset to make reliable predictions from representative data is critical, data collection strategies need to be chosen strategically since relying solely on keyword-based crawling can bring in noise and irrelevant\cite{Bhattacharya2017} data from a different context into the dataset. Therefore, a suitable filtering mechanism is essential for better quality data with high recall as well as precision. A semantic filtering mechanism \cite{Sheth2016,Phillips2017} as in the Twitris platform, can be implemented that selects and employs a domain-specific knowledge graph (e.g., using the Drug Abuse Ontology for related opioid analysis \cite{Cameron2013}) for precision, and reuses a broad knowledge graph such as DBPedia for coverage and generality (see section 2). Thus, a significant and relevant dataset can be collected with high recall and precision that will allow one to obtain insights on the user engagement.

\section{Applications}
Twitter data has enabled researchers and analysts to deal with diverse domains ranging from healthcare, finance, and economy to socio-political issues and crisis management. Approaches to retrieve as much information as possible requires the inclusion of domain-specific features as well as the use of domain knowledge in the analysis. In this section, we provide a list of domains where predictive analysis applications on Twitter were implemented, along with the technical details. A comprehensive table is also included at the end to give a comparative overview of application domains, the features and machine learning algorithms being used, and their performance. The included applications were selected because they were state-of-the-art in their respective domains or had been influential. The applications that we describe in this section combine a variety of signals that can be the basis for coarse-grained predictive analysis. Since some of the applications in this section make use of the Twitris platform; therefore, we first provide background information about the platform. Purohit et al. \cite{Purohit2013} introduced the Twitris platform for citizen sensing that performs analysis of tweets, complemented by shared information from contextually relevant Web of Data and background knowledge. They describe it as a scalable and interactive platform which continuously collects, integrates, and analyzes tweets to give more profound insights. They demonstrate the capabilities of the platform with an analysis in various dimensions including spatio-temporal-thematic, people-content network, and sentiment-emotion-subjectivity, with examples from business intelligence including brand tracking, advertising campaigns, social/political unrests, and disaster events.

\subsection{Healthcare}
Twitter data can be employed to shed light on many healthcare and disease-related aspects of contemporary interest, ranging from Alzheimer and dementia progression \cite{Robillard2013a} to eating disorders \cite{Prieto2014a} and mental health problems \cite{Yazdavar2017,Coppersmith2015}. We focus on applications to glean depression in individuals or at a community level using self-reports about these conditions, their consequences, and patient experiences on Twitter.

Depression is a condition that a sizable population in all walks of life experiences in their daily life. Social media platforms including Twitter has been used to voluntarily express the mood changes and feelings as they arise. From these tweets, it is possible to predict whether a user is depressed or not, what symptoms they show as well as the reasons for their depressive mood. Some examples indicative of depression as expressed in tweets\footnote{These tweets were modified before we share them in this chapter.} include: "I live such a pathetic life.", "Cross the line if you feel insecure about every aspect of your life." ,"That's how depression hits. You wake up one morning afraid that you're going to live.", and "Secretly having a mental breakdown because nothing is going right and all motivation is lost.". These tweets epitomize the expression of emotional tumult that may underlie subsequent conscious actions in the physical world.

An interesting study by Yazdavar et al. \cite{Yazdavar2017} explored the detection of clinical depression from tweets by mimicking the PHQ-9 questionnaire which clinicians administer to detect depression in patients. This study is different from traditional clinical studies that use questionnaires and self-reported surveys. They crawled 23M tweets over 45K twitter users to uncover nine significant depressive symptoms; (1) Lack of Interest, (2) Feeling Down, (3) Sleep Disorder, (4) Lack of Energy, (5) Eating Disorder, (6) Low Self-esteem, (7) Concentration Problems, (8) Hyper/Lower Activity, and (9) Suicidal Thoughts. A probabilistic topic model with a semi-supervised approach is developed to assess clinical depression symptoms. This hybrid approach is semi-supervised in that it exploits a  lexicon of depression symptoms as background information (top-down) and combines it with generative model gleaned from the social media data (bottom-up) to achieve a precision of 72\% on unstructured text. 

De Choudhury et al. \cite{DeChoudhury2013a} predicted the depression in an individual by exploiting their tweets. For ground truth dataset, they used, crowdsourcing to collect and label data. They utilized tweet metadata, network, statistical, textual and linguistic features, and time series analysis over a year of data to train an SVM model, obtaining an accuracy of 0.72.

The extraction of the location of people who experience depression using textual and network features can further assist in locating depression help centers. \cite{Do2017} utilizes a multiview\footnote{\url{ http://www.wcci2016.org/document/tutorials/ijcnn8.pdf}} and deep learning based model, to predict the user location. The multi-entry neural network architecture (MENET) developed for location prediction uses words, the semantics of the paragraph (using doc2vec \cite{Lau2016}), network features and topology (using node2vec \cite{Grover2016}) and time-stamps to deduce user’s location.  They achieved an accuracy over 60\% for GeoText\footnote{\url{ https://www.cs.cmu.edu/~ark/GeoText/README.txt}}, UTGeo11\footnote{\url{http://www.cs.utexas.edu/~roller/research/kd/corpus/README.txt}} and 55\% for TwitterWorld\cite{Bo2012}.  Furthermore, MENET achieves an accuracy of 76\% in region classification and 64.4\% in state classification using GeoText dataset. 

\subsection{Public Health}
Social media platforms including web forums, Reddit and Twitter, has become a venue where people seek advice and provide feedback for problems concerning public health. These conversations can be leveraged to predict trends in health-related issues that may threaten the well-being of the society. Moreover, caregivers have also seen these sources to be a game changer in its potential for actionable insights because of the information circulation. Particularly, cannabis legalization issue in the U.S. has been a trending topic\footnote{\url{http://www.pewresearch.org/fact-tank/2018/01/05/americans-support-marijuana-legalization/ft_18-01-05_marijuana_line_update/}} in the country as well as social media. Prior research on Twitter data analysis in this domain proved that it is an essential tool for epidemiological prediction of emerging trends. 

Existing studies have involved identifying syntactic and statistical features for public health informatics, such as PREDOSE (PRescription Drug abuse Online Surveillance and Epidemiology) which is a semantic web platform that uses the web of data, background domain knowledge and manually created drug abuse ontology for extraction of contextual information from unstructured social media content.  PREDOSE performs lexical, pattern recognition (e.g., slang term identification), trend analysis, triple extraction (subject-predicate-object) and content analysis. It is helpful in detecting substance abuse involving marijuana and related products.  Not only can it analyze generic marijuana but also its concentrates like butane hash oil, dabs, and earwax that are used in the form of vaporizers or inhalers. In a similar analysis of Twitter data, the marijuana concentrate use and its trends were identified in states where cannabis was legalized as well as not legalized. In 2014, utilizing the eDrugTrends\footnote{\url{http://wiki.knoesis.org/index.php/EDrugTrends}} Twitris platform, researchers collected a total of 125,255 tweets for a two-month period, and 22\% of these tweets have state-level location information\cite{Daniulaityte2015}. They found that the percentage of dabs-related tweets was highest in states that allowed recreational or medicinal cannabis use and lowest in states that have not passed medical cannabis laws, where the differences were statistically significant. A similar study in 2015 \cite{Lamy2016} reported adverse effects of Cannabis edibles and estimated the relationship between edibles-related tweeting activity and local cannabis legislation. Another study \cite{Daniulaityte2015} was to automatically classify drug-related tweets by user type and the source of communication as to what type of user has authored the tweet, where the user types are defined as user, retailer and media. They employed supervised machine learning techniques incorporating the sentiment of tweets (e.g., positive, negative, neutral). 

\subsection{Political Issues}
Political discussions on Twitter, which capture dynamic evolvement of public opinion, can directly impact the outcome of any political process. Arab Spring demonstrations \cite{Howard2011,Tufekci2014,Arpinar2016} in the middle eastern countries, Gezi protests \cite{Haciyakupoglu2015,Tufekci2014} in Turkey, as well as US Presidential elections in 2016 involving influence peddling on several social media platforms \cite{Allcott2017} provide impactful illustrative examples. Researchers have explored user classification and profiling in the context of such political events on Twitter to predict the issue trends and eventual outcome.

Researchers \cite{Pennacchiotti2011,Hoang2013a,Makazhanov2014} used Twitter data to predict political opinions of users based on linguistic characteristics (e.g., Tf-IDF) of their tweet content. While classification of users based on their political stance on Twitter has been well studied, Cohen et al. \cite{Cohen2013} have claimed that much of the studies and their datasets to date have covered very narrow portion of the Twittersphere, and their approaches were not transferable to other datasets. Pennacchiotti et al. \cite{Pennacchiotti2011} focused on the user profiling task on Twitter, and used user-centric features such as profile, linguistic, behavioral, social and statistical information, to detect their political leanings, ethnicity, and affinity for a particular business.

Moreover, prediction of dynamic groups of users has been employed \cite{Chen2012} to monitor the polarity during a political event by analyzing tweeting behavior and content through clustering. Usage of hashtags and URL, retweeting behaviors and semantic associations between different events were key to clustering. 56\% of the Twitter users participated in 2012 US Republican Primaries by posting at least one tweet, while 8\% of the users tweeted more than 10 tweets. 35\% of all users mostly retweet, separating them from the remaining. In terms of dynamic user groups, they formed the following bilateral groups: silent majority and vocal minority, high and low engaged users, right and left-leaning users, where users were from different political beliefs and ages. They analyzed these dynamic groups of users to predict the election outcomes of Super Tuesday primaries in 10 states. They also reported that the characterization of users by tweet properties (frequency of engagement, tweet mode, and type of content) and political preference provided insights to make reliable predictions. 8 weeks of data comprising 6,008,062 tweets from 933,343 users about 4 Republican candidates: Newt Gingrich, Ron Paul, Mitt Romney and Rick Santorum, was analyzed to assess the accuracy of predicting the winner. Prediction of user location using a knowledge base such as LinkedGeoData\footnote{\url{http://linkedgeodata.org/About}} in tweets also contributed to the election prediction.  Furthermore, an error of 0.1 between the prediction and actual votes attest to the efficacy of the approach.  Such a low error rate in prediction is attributed to original tweets (not retweets)  from users who are highly engaged and right leaned. 

\subsection{Social Issues}
Social issues and related events have been a part of discussions on Twitter, which gives opportunities to the researchers to address problems concerning individuals as well as the society at large. Solutions to such problems can be provided by measuring public opinion and identification of cues for detrimental behavior on Twitter by employing predictive analysis. We explain three problems and their respective solutions in this subsection.

\subsubsection{Harassment}
Harassment\footnote{\url{http://wiki.knoesis.org/index.php/Context-Aware_Harassment_Detection_on_Social_Media}} is defined as an act of bullying an individual through aggressive offensive word exchanges leading to emotional distress, withdrawal from social media and then life. According to a survey from Pew Research Center\footnote{\url{http://www.pewinternet.org/2014/10/22/online-harassment/}}, 73\% of the adult internet users have observed, and 40\% have experienced harassment, where 66\% percent of them are attributed to social media platforms. Also, according to a report from Cyberbullying\footnote{\url{http://cyberbullying.us/facts}} research center, 25\% of teenagers claimed to be humiliated online. While it is imperative to solve this problem, frequency and severe repercussions of online harassment exhibit social and technological challenges. 

Prior work \cite{Xu2012} has modeled harassment on social media to identify the harassing content which was a binary classification approach. However, in their predictive analysis, the context, network of users and dynamically evolving communities shed more light on the activity than pure content-based analysis. For instance, sarcastic communication between two friends on social media may not be conceived as harassment while the aggressive conversation between two strangers can be considered as an example of bullying. For reliably identifying and predicting harassment on Twitter, it is essential to detect language-oriented features (e.g., negation, offensive words), emotions, and intent. \cite{Chen2012b} employs machine learning algorithms along with word embedding, and DBpedia knowledge graph to capture the context of the tweets and user profiles for harassment prediction. 

Edupuganti et al. \cite{Edupuganti2017} focused on reliable detection of harassment on Twitter by better understanding the context in which a pair of users is exchanging messages, thereby improving precision. Specifically, it uses a comprehensive set of features involving content, profiles of users exchanging messages, and the sequence of messages, we call conversation. By analyzing the conversation between users and features such as change of behavior during their conversation, length of conversation and frequency of curse words, the harassment prediction can be significantly improved over merely using content features and user profile information. Experimental results demonstrate that the comprehensive set of features used in our supervised machine learning classifier achieves F-score of 88.2 and Receiver Operating Characteristic (ROC) of 94.3.
Kandakatla et al. \cite{Kandakatla2016} presents a system that identifies offensive videos on YouTube by characterizing features that can be used to predict offensive videos efficiently and reliably. It exploits using content and metadata available for each YouTube video such as comments, title, description, and the number of views to develop Naïve Bayes and Support Vector Machine classifiers.The training dataset of 300 videos and test dataset of 86 videos were collected, and the classifier obtained an F-Score of 0.86.

\subsubsection{Gang Communities \& Their Members and Gun Violence}
Gang communities and their members have been using Twitter to subdue their rivals, and identification of such users on Twitter facilitates the law enforcement agencies to anticipate the crime before it can happen. Balasuriya et al. \cite{Balasuriya2016} investigated conversations for finding street gang members on Twitter. A review of the profiles of gang members segregates them from rest of the Twitter population by checking hashtags, YouTube links, and emojis in their content \cite{Wijeratne2015}. In \cite{Balasuriya2016}, nearly 400 gang member profiles were manually identified using seed terms, including gang affiliated rappers, their retweeters, followers as well as followees. They used textual features of the tweet, YouTube video descriptions and comments, emojis and profile pictures to power various machine learning algorithms including Naive Bayes, Logistic Regression, Random Forest and Support Vector Machines, to train the model. Random Forest performed well for Gang and Non-Gang classification. It is interesting to notice that gang members usually make use of their profile images in a specific way to intimidate other people and members of rival gangs. 

As gun control policies in big cities, such as Chicago, have changed over the years, the volume of the taunting and threatening conversations on social media has also relatively increased \cite{Blevins2016}. Such conversations can be leveraged to assist law enforcement officers by providing insights on situational awareness as well as predicting a conflict between gang groups for a possible gun violence incident. Blevins et al. \cite{Blevins2016} used a Twitter dataset that was manually labeled by a team of researchers with expertise in cyber-bullying, urban-based youth violence and qualitative studies. Their strategy was to collect all tweets, mentions, replies, and retweets from a specific user profile between 29 March and 17 April 2014. Three experts developed the key types of content and used the work by Bushman and Huesmann \cite{Bushman2006} to identify and categorize types of aggression. To overcome the challenge of recognizing special slang terms and local jargons in tweets as mentioned in Desmond et al. work, Blevins et al.\cite{Blevins2016} developed a part-of-speech (POS) tagger for the gang jargon and mapped the vocabulary they use to Standard English using machine translation alignment. They developed emotion classifier that uses the extracted POS tags, and Dictionary of Affect in Language (DAL) quantitative scores (Whissell, 2009) as key features.  Ternary classification is applied to the whole dataset (TCF) and binary classification on the aggression-loss subset (BCS). Then they use a  cascading classifier (CC), which uses two SVM models. Initially, one SVM model is used to filter the tweets into aggression/loss tweets, and all other tweets fall into the other category. After this filtration, only aggression/loss tweets is passed to second SVM model which is again a binary classifier for loss or aggression. So this Aggression Supervised classifier is able to categorize loss with 62.3\% F-score and aggression with 63.6\% F-score which beats the baseline model (Unigrams) by 13.7 points (aggression) and 5.8 points (loss) \cite{Blevins2016}. 

\subsection{Transportation}
Congestion due to traffic is one of the prevalent problems in the United States (U.S.). Even after having structured rules that govern the flow of the traffic in the U.S., congestion due to non-recurring activities still affects the schedules of people. However, the stationing of police officers to smooth the traffic is a probable solution, although it would not be long-term. Having an estimation of the flow of traffic in the advent of an event can help people to re-route their path to the destination. Leveraging social media and machine learning to estimate traffic is one such long-term solution that can be drafted for active traffic monitoring. Social media is flooded with posts from people about an event. Such posts can provide the location of the event or the tweeter, and it can be used along with other textual features to estimate the traffic flow. In \cite{Ni2014}, textual features, tweet and user-metadata such as text, hashtags, URLs, number of users and retweeted tweets were used by combining with live event data to predict traffic dynamics. They utilized autoregressive model, neural network, support vector regression, and K-nearest neighbor for traffic prediction. The evaluation was performed using mean absolute percentage error (MAPE), and root means square error (RMSE), with support vector regression (SVR) performing better over other regression models. SVR reduced the error in traffic prediction by 24\% in terms of RMSE. 

\subsection{Location Estimation}
Social media serves a vital role in times when people struggle to survive a disastrous event such as hurricane or earthquake, to provide solutions for assisting the public in recovery efforts. These solutions include identification of the demand and its location, and mapping the identified demands with suitable suppliers analyzing Twitter data.

In particular, location extraction plays a significant role in identifying the area that is impacted by a disaster as well as providing assistance \cite{Krishnamurthy2014}. Mahmud et al. \cite{Mahmud2012a} developed an approach to predict the location of users at the city level on Twitter combining several classifiers. They removed stop words, performed part-of-speech tagging, extracted hashtags, and extracted a feature called local term,  a term used by local people to refer to the city. For detecting the local terms, several classification algorithms and found Naïve Bayes, SVM and Decision trees (J48) as the best performing algorithms. Al-Olimat et al. \cite{Al-Olimat2017} developed a tool called LNEx (Location Name Extraction), that extracts the location from the tweet content by utilizing the OpenStreetMap\cite{Haklay2008}, GeoNames \cite{Ahlers2013} and DBpedia \cite{Lehmann2012} for disambiguation. The information retrieval process from the tweet is two-fold, which are toponym extraction and geoparsing. Toponym is a process to extract city and street names, points of interest, from unstructured text, tweets in particular for this study. Location names are usually abbreviated on Twitter; hence, a text normalization procedure is used for expansion of such brevity. For instance, tweets may contain “Rd” as an abbreviation, and it is normalized to “road”. Furthermore, ambiguous location problems are resolved by employing the geoparsing procedure using the OpenStreetMap API\footnote{\url{https://wiki.openstreetmap.org/wiki/API_v0.6}}. LNEx improved the average F-Score by 98-145\%, outperforming all the state of art taggers.

\subsection{Community on Social Media}
People with distinct feelings, expression, solutions, and intelligence, share their opinions on Twitter.  Such a diversified content can be related to elections, football game or a domain that is influenced by public views. With the abundance of textual data, one can envision the power of collective intelligence that can be harnessed for a wise recommendation, judgment and strategy building. Also, it is a known fact that a judgment call made by a crowd is superior to an individual’s decision \cite{Lee2017}. Formation of a diverse group can improve the decision-making process through what is known as Wisdom of Crowd (WoC). WoC is meant to minimize regional biases that may cloud objectivity associated with individual’s judgment and bring together different perspectives and knowledge that can enhance coverage and comprehensiveness of the analysis. For example, WoC can be used to design a portfolio of stocks that maximize the profit in the stock market trading. However, no existing work illustrates the notion of WoC statistically and analytically. A methodological way for measuring the diversity of the crowd is crucial to the rise of human social engagement on social media. According to a recent survey from Pew Research Center, 76\% of the American population is active on social media. It attributes success to a significant amount of online data and can aid in creating WoC of the social system. In \cite{Bhatt2017}, fantasy premier league (FPL) is considered to exercise the better judgment of the diverse crowd. In their work, they predict the best performing team captain in the premier league, an element dictating the success of a team, based on the scores retrieved from the fantasy football and content of Twitter users. They utilized Word2vec similarity measure to quantify the diversity of two groups of users during captain selection in FPL. Furthermore, They defined and validated their statistical objective scoring criteria to measure the quality of crowd judgment. 

\subsection{Demographics} 
In many applications, demographic information is a key to analysis that depends on different segments of the population concerning age groups, ethnicity, and gender. For example, age is critical for understanding drug abuse, while gender is critical to understand vulnerability to depression. Twitter in its current state does not require users to provide any demographic information.

\subsubsection{Age Estimation using social media}
Researchers developed a machine learning system coupled with the DBpedia knowledge graph utilizing the user follower-followee networks to predict the most probable age of a Twitter user, in \cite{Smith2018}. They gathered pre-identified famous people from DBpedia, based on their occupations and areas of interest, which also included their birth dates. Then they extracted a sample of 23,120 users who are in one/two hops of follower-followee network of famous people. Some of the user profiles were spam/bot and hence they were removed. Then they selected 16K users among the followers of the top 50 famous figures as their training set and 8K as their testing set. They achieved 84\% accuracy in predicting the age of these users. They selected Support Vector Regression (SVR) with K-Fold Cross Validation \cite{Refaeilzadeh2009} as their best performing model after evaluating using Linear Regression \cite{Nguyen2011}, Least Absolute Shrinkage and Selection Operator (LASSO) \cite{Chen2010Face}, and ElasticNet \cite{Culotta2016}.

Zhang et al. \cite{Zhang2016a} studied the problem of age prediction on Twitter, using SVM and least square optimization algorithm in building the model. They utilized various features such as linguistic, textual, and network, to improve their model, achieving an F1 score of 0.81. They discovered that the characteristics of users in the same age groups have similar content and interactions between each other. On the other hand, Nguyen et al. \cite{Nguyen2013a} investigated the relationship between the language used in tweets of a user and his/her age. They annotated the dataset that was collected following a guideline formed based on the tweet content of users in different age groups such as explicit or implicit age or life stage mentions. They found that the language use of people in same age groups is similar regarding the word and phrase selections as well as the topics that they are talking about. For instance, the following two sets of words, “school, son, daughter, wish, enjoy, thanks, take care” and “haha, xd, internship, school” have been used by users in two different age groups. In their analysis, they used linear and logistic regression models with unigram feature only, achieving an F1 score of 0.76.

\subsubsection{Gender Estimation using Twitter}
Estimation of the gender of a twitter user is beneficial to the analysis of Twitter data for health-related, drug abuse, and harassment activities. Existing approaches utilized statistical features \cite{Bamman2012} and seldom involved background knowledge along with social information. In \cite{Li2017a}, a dataset from Sina Weibo, which is a counterpart of the micro-blogging platform Twitter, in China, was used to assess their methodology for gender prediction. \cite{Li2014} exploits online behavioral and textual features and choice of vocabulary for each user.  Online behavioral features include the number of fans, attention, messages, comments, forwards and a ratio of original/forwarded messages. Textual features include hashtags, URLs, emoticons, and sentence length. They also made use of username and pictures in content. Lexical features were extracted from the content using TF-IDF.  They used four algorithms for predicting gender: Decision Tree, Naive Bayes, Logistic Regression and Support Vector Machines (SVMs), and found that SVM outperformed other classifiers by attaining accuracy of 94.3\%.  

\subsection{Anomaly \& Popularity Prediction}
Twitter has become a playground for spammers. While public conversations on Twitter are diverse and challenging to analyze and summarize, spammers and bots further complicate the reliability of the outcome. Bots are automated software that is programmed to post a predefined content. They are being used mostly to propagate or promote bias and skew votes in politics, views on social issues, or provide impetus to promotional campaigns. On the other hand, prediction of the popularity of trending topics or issues requires robust analysis that takes into account anomalous accounts.

Thomas et al. \cite{Thomas2011a} collected 1.8 billion tweets sent by 32.9 million users and manually identified 1.1 million suspended accounts as spammer accounts along with 80 million anomalous tweets. They used user behavior regarding interactions with other users, public Twitter handler service usage and textual features of tweet content such as shortened URLs created using free web hosting services. Volkova et al. \cite{Volkova2017a} also studied this problem by applying a deep learning technique, Recurrent Neural Networks (RNN), using tweet metadata and network features. They compared their approach with state-of-the-art machine learning methods such as log-linear models. Their RNN model outperformed all the machine learning models built using various combinations of features with 0.95 F1 score. Sentiment has also been used in spam detection works \cite{Dickerson2014,Varol2017a} as a feature to detect bots on Twitter. Varol et al. \cite{Varol2017a} also studied the detection of online bots on Twitter, and utilized Random Forests, AdaBoost, Logistic Regression and Decision Tree algorithms. They found Random Forest classifier achieved the best performance with 0.95 AUC score. They made use of sentiment features that they extracted from the text beside tweet and user metadata, textual, linguistic and network features.

Poblete et al. \cite{Poblete2011a} have investigated the tweet credibility issue in the news disseminated on the platform. They crowdsourced the task of evaluating the credibility of each tweet to determine if it has newsworthy topics, labeling each tweet using automated credibility analysis. Labels given by crowdsourcing process were used in the training phase. They used SVM, decision trees, decision rules and Bayesian networks, and best results were given by J48 decision tree, achieving an 86\% F1 score. Ross et al. \cite{Ross2016} created a robust and general feature set for learning to rank tweets based on credibility and newsworthiness. In previous works by Gupta et al. \cite{Gupta2014TweetCred,Gupta2012,Gupta2013}, they have demonstrated that when the training and testing data are from two distinct time periods, the ranker performs poorly. Ross et al. \cite{Ross2016} improved upon this by creating a feature set that does not overfit a particular year or a set of topics, which is critical for robust analysis of social media over time and across different domains. \\
Varol et al. \cite{Varol2017b} conducted a time series analysis to predict if a trending meme is organic or promoted by a group. They aimed to predict meme’s that have potential to trend before it becomes trending; therefore, the task of predicting trends is naturally forced to utilize a sparse dataset. For this reason, they had to reliably extract textual, linguistic, tweet and user metadata, network and statistical features, from a small dataset.  They used three learning algorithms namely, K-Nearest Neighbor (KNN) with Dynamic Time Warping (KNN-DTW), Symbolic Aggregate approXimation with Vector Space Model (SAX-VSM) and KNN. KNN is a machine learning algorithm for classification and DTW for multi-dimensional time series.  They found KNN-DTW and KNN showed the best performance in prediction. They used AUC as evaluation metric to measure accuracy because it is not biased by the imbalance in classes (e.g., 75 promoted trends versus 852 organic ones). Weng et al. \cite{Weng2014a} studied the prediction of the popularity of meme on Twitter. They relied mostly on network features besides tweet and user metadata, using random forest and linear regression. They extracted 13 features such as some early adopters, average shortest network path length between users, the diameter between users, and the number of infected communities. They built their model using random forest and tested against five different baselines that used linear regression along with different combinations of the 13 features.  Their model achieved 0.85 F1 score, outperforming the baselines. Kobayashi et al. \cite{Kobayashi2016a} predicted the popularity of a tweet in terms of the number of retweets in a time window in the future. They used time series analysis using a method called time-dependent Hawkes process (TiDeH) calculating infectious rate and using tweet and user metadata such as temporal information from a tweet and number of followers of a user. They evaluated their system against other existing methods that incorporated linear regression and Poisson process and reported that it outperformed other approaches achieving around 5\% mean error rate. Tsur et al. \cite{Tsur2015a} also studied the popularity of hashtags on Twitter, through linguistic features of the tweet text, specifically hashtags. They obtained promising results using a modified version of Gradient Boosted Trees called Gradient Boosted rank. They compared their approach with SVM and Least-effort algorithms, obtaining 0.11 mean error rate. Ruan et al. \cite{Ruan2012a} predicted the volume of tweets, analyzing the user behavior on individual as well as collective level. Besides tweeting activity and content analysis of users, they utilized the underlying follower-followee network, user network structure, neighboring friends’ influence and user past activity as features. They used linear regression model with multiple features that include network structure, user interaction, content characteristics and past activity, and found that combining features yields the best performance.

\begin{table*}[!htbp]
\centering
\caption{ Comparative Analysis of Applications and their Evaluation. 
Acronyms for Algorithms and Features are described in Table \ref{table_2}}
\label{table_1}
\scriptsize
\begin{tabular}{@{\extracolsep{\fill}}|p{3cm}|p{4cm}|p{3cm}|p{3cm}|}\toprule
\textbf{Ref. \& Evaluation}                        & \textbf{Application}                                                                        & \textbf{Algorithms  }                                                                                                                                               & \textbf{Features } \\
\hline
  &   &  &               \\
\cite{Pennacchiotti2011} $\>$ $\>$ F1=0.88    &   & SVM & UsM, TwM                                                   \\
\cite{Zhang2016a} $\>$ F1=0.81  &  User Profiling  & LinR & Ling, Nw,                                                                                                                              \\
\cite{Pattisapu2017} $\>$ F1=0.59      &    & CNN  &  Stat, Txt                                                                                                                           \\
\cite{Gilani2017a} $\>$ F1=0.83      &  User Classification  & RF  & Vis                                                                                                                            \\
\cite{Balasuriya2016} $\>$ $\>$ F1=0.77    &   &   NB    &                                                                                                                              \\
\cite{Alowibdi2014} $\>$ Acc=0.82                  &                                                                                       &         LogR, LASSO                                                                                                                                         &                                                                                                                              \\
\cite{Hoang2013a} $\>$ $\>$ Acc=0.92                    &                                                                                       &                                                                                                                                                            &                                                                                                                              \\
\cite{Makazhanov2014} $\>$ $\>$ F1=$\sim$0.75           &                                                                                       &                                                                                                                                                            &                                                                                                                              \\
\cite{Nguyen2013a} $\>$ F1=0.76                    &                                                                                       &                                                                                                                                                            &                                                                                                                              \\
\cite{Lewenberg2015} $\>$ $\>$ AUC=$\sim$0.7 ﾊ          &                                                                                       &                                                                                                                                                            &                                                                                                                              \\
\cite{Wagner2013a} $\>$ $\>$ AUC=0.8                    &                                                                                       &                                                                                                                                                            &                                                                                                                              \\
\cite{Smith2018} $\>$ Acc=0.84                     &                                                                                       &                                                                                                                                                            &                                                                                                                              \\
 & & &   \\
\hline
 & & &   \\
\cite{DeChoudhury2013a} $\>$ $\>$ Acc=0.72              & User Attitude, Personality,                                          & SVM, NB, RF                                                                                                                            & TwM, Txt, Ling,                                                                     \\
\cite{Mahmud2016} $\>$ $\>$ F1=0.62                     & and Mood Prediction                                                                                    &                                                                                                                                                            &  Nw, Stat                                                                                                                              \\
 & & &   \\
\hline
 & & &   \\
\cite{Georgiev2014a} $\>$ AUC=0.8                  & Sales \& Stock price prediction                                              & NB, RF, SVM                                                                                                                             & TwM                                                                                                               \\
  & & &   \\
\hline
 & & &   \\
\cite{DeChoudhury2016} $\>$ $\>$ Med error=0.32         & Social and Political events,                            & PosR  & TwM, UsM                                                                       \\
\cite{Yang2017}{]} $\>$ F1=0.58              &   Elections, Collective action                                                                                      
& NBR, SVM,                                                                                                                                                         & Txt, Ling, Nw                                                                                                                            \\
\cite{Korolov2016} $\>$ Acc=0.85                   &                                                                                       &           LogR, CNN, RF                                                                                                                                                   &                                                                                                                              \\
\cite{Kallus2014} $\>$ AUC=0.91                    &                                                                                       &                                                                                                                                                            &                                                                                                                              \\
 & & &   \\
\hline
 & & &   \\
\cite{Varol2017b} $\>$ AUC=0.95                 &                                                                  & KNN-DTW, SAX-VSM, & Txt                                                      \\
\cite{Weng2014a} $\>$ F1=$\sim$0.85          & Popularity prediction                                                                                      &  KNN, RF,                                                                                                                                                           & Nw and Stat,                                                                                                                           \\
\cite{Kobayashi2016a} Mean err= $\sim$0.179 &                                                                                       &  TiDeH, LinR, LogR                                                                                          & TwM and UsM                                                                                                                                                                      \\
\cite{Tsur2015a} Mean err=$\sim$0.11        &                                                                                       &   GrB, SVM                                                                                                                                                       &   Ling                                                                                                                              \\
 & & &   \\
\hline
 & & &   \\
\cite{Thomas2011a} $\>$ Acc=0.89                   &                                     & KNN-DTW, SAX-VSM, & Text, Nw,                                                     \\
\cite{Varol2017b} $\>$ AUC=0.95                 &   Spam bot detection                                                                                    
&       KNN, LinR, DT                                                                    
&       Ling, TwM, UsM                                                                                                                      \\
\cite{Dickerson2014} $\>$ AUC=0.73                 &                                                                                        &  NB, GrB                                                                                                                                    
&  Stat, Vis                                                                                                                           \\
\cite{Volkova2017a} $\>$ F1= 0.95                  & Troll detection                                                                                     &  RF, AdB,                                                                                                                                          &                                                                                                                              \\
\cite{Echeverria2017} $\>$ F1=0.99                 &                                                                                       & BNet, RNN,                                                                                                                                                            &                                                                                                                              \\
\cite{Varol2017a} $\>$ AUC=0.95                 & Credibility prediction                                                                                     & SVM, LogR                                                                                                                                                          &                                                                                                                              \\
\cite{Poblete2011a} $\>$ F1=0.86                   &                                                                                       &                                                                                                                                                            &                                                                                                                              \\
 & & &   \\
\hline
 & & &   \\
\cite{Davidov2010} $\>$ Pre=$\sim$0.80             & Sentiment analysis                          & NB, SVM, CRF,                                                     & Ling,                                                                                 \\
\cite{Wang2011} $\>$ F1=0.77                 &                                                                                      &         LibLin, RBF-NN, AdB,                                                                                                                                                    &    Txt,                                                                                                                          \\
\cite{Kouloumpis2011a} $\>$ F1=0.67                & Emotion Detection                                                                                       &  RT, REPTree,                                                                                                                                                        &  TwM                                                                                                                              \\
\cite{Agarwal2011} $\>$ F1=$\sim$0.60          &                                                                                       &       BNet, LogR                                                                                                                                                        &        and UsM                                                                                                                      \\
\cite{Pak2010} $\>$ F0.5=0.62                      &                                                                                       &                                                                                                                                                            &                                                                                                                              \\
\cite{Go2009} $\>$ Acc=0.82                        &                                                                                       &                                                                                                                                                            &                                                                                                                              \\
\cite{Wang2012} $\>$ $\>$ Acc=0.61                      &                                                                                       &                                                                                                                                                            &                                                                                                                              \\
\cite{Gao2015} $\>$ Mean error=0.071               &                                                                                       &                                                                                                                                                            &                                                                                                                              \\
\cite{Hassan2013} $\>$ Acc=$\sim$0.71            &                                                                                       &                                                                                                                                                            &                                                                                                                              \\
\cite{Kothari2013} $\>$ F0.5= 0.76                 &                                                                                       &                                                                                                                                                            &                                                                                                                              \\
\cite{Nguyen2012} $\>$ $\>$ F1=$\sim$0.73               &                                                                                       &                                                                                                                                                            &                                                                                                                              \\
\cite{Liu2012} $\>$ $\>$ F1=0.79                        &                                                                                       &                                                                                                                                                            &                                                                                                                              \\
 & & &   \\
\hline
 & & &   \\
\cite{Mahmud2012a} $\>$ $\>$ Acc=$\sim$0.83             & Location Estimation                                               & NB, SVM, DT,                                                                                                & Ling, TwM, UsM                                                                                \\
\cite{Georgiou2015a} $\>$ Mean error=0.39          &                                                                                         &  LinR, MxEnt                                                                                                                                                             &  Txt                                                                                                                          \\
\cite{Aiswal2013} $\>$ Acc=0.88                    & Traffic Estimation                                                                                      &                                                                                                                                                              &                                                                                                                            \\
\cite{Rout2013} $\>$ Acc=0.79                      &                                                                                       &                                                                                                                                                            &                                                                                                                                \\
 & & &   \\
\hline
 & & &   \\
\cite{Rath2017} $\>$ F1= 0.70                      & Finding Important Users,                                          & RNN, SVM                                                                                                                                                   & Nw, UsM, Txt                                                                                              \\
\cite{Bizid2015} $\>$ Pre=0.86                     & Community Detection                                                                                       &                                                                                                                                                            &                                                                                                                              \\
 & & &   \\
\hline
 & & &   \\
\cite{Ferrara2013} $\>$ LFK-NMI=0.13               & Topic Extraction,                                                   & HiCl, KM, LDA, SVM                                                                                                                 & Txt, Ling, TwM, UsM                                                                         \\
\cite{Yamamoto2015} $\>$ F1=0.63                   &  Meme Extraction                                                                                        &                                                                                                                                                            & Nw                                                                                                                             \\
 & & &   \\
\hline
 & & &   \\
\cite{Beykikhoshk2014} $\>$ Acc=0.84               & Public Health  & NB,                                   & Txt, Ling, TwM \\
\cite{Yin2016} $\>$ AUC=0.83  &  	&	 SVM, RF                                                                                                                                        &    Sentiment,                                                                                                                         \\
\cite{Daniulaityte2016} $\>$ F= 0.8736             &  Health-care                                                                                      &  LDA, ssToT                                                                                                                                                          &                                                                                                                                \\
\cite{Lamy2016} $\>$ $\>$ K Alpha=0.84                  &                                                                                      &                                                                                                                                                             &                                                                                                                               \\
 & & &   \\
\hline
 & & &   \\
\cite{Al-Olimat2017} $\>$ $\>$  F1=0.81                  & Disaster Management                                                                   & LogR                                                                                                                                                           & Txt, Ling, UsM, TwM                                      \\
\cite{Hu2015} $\>$ R2= 0.67                        &                                                                                       &                                                                                                                                                            &  Nw    \\                                                                                                           
 & & &   \\
\hline
\end{tabular}
\end{table*}

\subsection{Sales \& Stock Price Prediction}
As social media, particularly Twitter, users share their satisfaction or frustration with products on the platform, these user reviews can be exploited by companies to generate actionable insights to meet customer expectations and eventually provide better quality products and services. Industrial applications of predictive analysis of social media have been gradually adopted, to gain the understanding of the market. Some of the use-cases that have adopted social media data for decision making are for:
\begin{enumerate}
	\item Improvement of Customer Service:  Delta Airline exploited social media to identify the reasons for customer frustration. For instance, lost luggage or poor service.
	\item New Products Research and Development: JD Power quality assessment has determined that car company modify car seats based user sentiments on the social sphere \cite{Kessler2010}.
	\item Key Influencers: A cosmetics company L'Oreal uses social media follower-followee network to find Influencers for promotions\footnote{https://www.socialmediatoday.com/special-columns/adhutchinson/2015-09-09/big-brand-theory-loreal-stays-connected-their-audience}.
    \item Recommendations through deep learning: YouTube utilizes the deep neural network to enhance their recommendation system using implicit feedback by analyzing users’ comments and videos of interest \cite{Covington2016}.
\end{enumerate}

Georgiev et al. \cite{Georgiev2014a} investigated the question of how the Olympic Games impacted the sales of businesses in London. They used Twitter posts along with the check-ins through Foursquare platform to extract mostly location-specific features from tweet text and tweet metadata, such as the distance of businesses to stadiums and sponsor businesses, transitions to entertainment places and social areas. They evaluated their work using AUC, for Naïve Bayes, Random Forest, and SVM algorithms and reported that SVM performed best with 0.8 of AUC.

\cite{Korpusik2016} employs feed-forward network (FF) for predicting the likelihood of a customer to buy a product. They restricted their dataset to tweets about mobile phones and cameras,  expensive products that people often buy after doing some research online. Before predicting the likelihood of purchasing a product, they predicted whether a tweet represents the respective user’s purchasing behavior. Then they predict whether the user will purchase the product after 60 days time window of tweeting. They compared the performance of their approach with Long Short Term Memory (LSTM), Recurrent Neural Networks (RNNs) (with varying dropout rates) based implementation and observed that their approach with FF surpasses others by small margins. FF learning cycle involved RMSprop \cite{Tieleman2012}, sigmoid activations and negative log-likelihood function with batch training.

\section{Conclusion}
Twitter has positioned itself as an essential part of the social media environment becoming an emergent communication medium. This development has opened up new opportunities for researchers to gauge the pulse of the populace reliably and use that to study public opinion, form policies, understand the impact of events, and find newer ways to address certain problems. Social media data has already enabled researchers to predict the trends and outcomes of several critical real-world events, and its reliability and coverage can further be improved by incorporating background knowledge \cite{Tufekci2014,Morstatter2013}. Specifically, monitoring the engagement and public opinion about ongoing events from temporal and spatial perspectives can foretell their evolution as well as the outcome. Moreover, this information can complement traditional surveys or polls that are conducted by non-government agencies to improve our confidence in the prediction, as traditional methods alone can be misleading or sluggish in reacting to rapidly changing events. In order to account for the complex decision making that requires consideration of a number of factors that can impact a situation or an event,  incorporation of as many signals as possible in comprehending the big picture is necessary. We have explored a predictive analysis paradigm that comprises two levels of prediction, using coarse-grained analysis built upon  fine-grained analysis.  Such analysis have been conducted with creditable success for events such as elections, gun violence, drug misuse or illicit drug use \cite{Sheth2018a}.

In this chapter, we have discussed processes, algorithms, and applications concerning predictive analysis in different domains. We illustrated fine-grained analysis by customizing domain-independent approaches to extract signals such as sentiment, emotions, and topics through the application of machine learning models, and coarse-grained analysis by aggregating and cultivating the signals to make predictions. We have also provided details of related prominent studies in ten different domains such as healthcare, public health, political and social issues, disaster management, sales and stock prediction, and demographics. The following table summarizes related work describing various applications and methods used.

\begin{table*}[!htbp]
\centering
\caption{ The Acronyms used in the comparative table.}
\label{table_2}
\scriptsize
\begin{tabular}{@{\extracolsep{\fill}}|p{1.5cm}|p{3cm}||p{1.5cm}|p{3cm}|}\toprule
\textbf{Acronym}    & \textbf{Algo. Description} & \textbf{Acronym} & \textbf{Feature Descr.} \\
\hline
LinR  &  Linear Regression & UsM & User metadata  \\
RF & Random Forest &TwM & Tweet metadata \\
NB  & Naïve Bayes & Ling & linguistic  \\
LogR  & Logistic Regression & Nw & Network  \\
PosR & Poisson Regression & Stat & Statistical \\
NBR & Negative Binomial Reg. & Txt & Textual \\
GB  & Gradient Boosting & Vis & Visual \\
AdB  & AdaBoost & & \\
DT  & Decision Trees  & &  \\
BNet  & Bayes Net  & &  \\
LibLin  & LIBLINEAR  & &  \\
HiCl    & Hierarchical Clustering  & &  \\
KM    & K-Means  & &  \\
RT    & Random Tree  & &  \\
\hline
\end{tabular}
\end{table*}

\section{Acknowledgement}
We are grateful to Amelie Gyrard, Mustafa Nural, Sanjaya Wijeratne, Shreyansh Bhatt, Ankita Saxena for their assistance with their reviews and comments that greatly improved this book chapter.

We acknowledge partial support from the National Science Foundation (NSF) award CNS-1513721: "Context-Aware Harassment Detection on Social Media", National Institutes of Health (NIH) award: MH105384-01A1: "Modeling Social Behavior for Healthcare Utilization in Depression", NSF award EAR- 1520870 “Hazards SEES: Social and Physical Sensing Enabled Decision Support for Disaster Management and Response”, Community in Social Media : This work was supported by Army Research Office Grant No. W911NF-16-1-0300, National Institute on Drug Abuse (NIDA) Grant No. 5R01DA039454-02 Trending: Social media analysis to monitor cannabis and synthetic cannabinoid use. Any opinions, conclusions or recommendations expressed in this material are those of the authors and do not necessarily reflect the views of the NSF, NIH, NIDA or Army Research Office.

\bibliography{Pred.bib}

\begin{thebibliography}{100}
\bibitem{Tieleman2012}
T.~Tieleman and G.~Hinton, ``Divide the gradient by a running average of its
  recent magnitude. coursera: Neural networks for machine learning. technical
  report.''
\end{thebibliography}
\bibliographystyle{IEEEtran}

\end{document}